\documentclass[aps,prc,twocolumn,groupedaddress]{revtex4}

\usepackage{amsmath}
\usepackage{amssymb}
\usepackage{graphicx}
\usepackage{bm}
\usepackage[dvips]{color}

\newcommand{\bftau}{\mbox{\boldmath $\tau$}}

\newcommand{\bfsigma}{\mbox{\boldmath $\sigma$}}

\def\lsim{\mathrel{\rlap{\lower4pt\hbox{\hskip1pt$\sim$}}
    \raise1pt\hbox{$<$}}}         

\def\gsim{\mathrel{\rlap{\lower4pt\hbox{\hskip1pt$\sim$}}
    \raise1pt\hbox{$>$}}}         

\begin{document}

\title{Microscopic description of the pygmy and giant electric dipole resonances \\
       in stable Ca isotopes.}

\author{ G.Tertychny$^{a,b}$, V.Tselyaev$^{a,c}$,
         S.Kamerdzhiev$^{a,b}$
         F.Gr\"ummer$^{a}$,
         S.Krewald$^{a}$, J.Speth$^{a}$,
         A. Avdeenkov$^{a,b}$,  E.Litvinova$^{b}$,\\
         {\it $^a$Institut f\"ur Kernphysik, Forschungszentrum J\"ulich,} \protect \\
         {\it 52425 J\"ulich, Germany} \protect \\
         {\it $^b$Institute of Physics and Power Engineering,} \protect \\
         {\it  249020 Obninsk, Russia}\\
         {\it $^c$Institute of Physics S.Petersburg University, Russia}}

\date{}

\begin{abstract}
The properties of the pygmy (PDR) and giant dipole resonance
(GDR)in the stable $^{40}Ca$,$^{44}Ca$ and $^{48}Ca$ isotopes have
been calculated within the \emph{Extended Theory of Finite Fermi
Systems}(ETFFS). This approach is based on the random phase
approximation (RPA) and includes the single particle continuum as
well as the coupling to low-lying collectives states which are
considered in a consistent microscopic way. For $^{44}Ca$ we also
include pairing correlations. We obtain good agreement with the
experimental data for the gross properties of both resonances. It
is demonstrated that the recently measured A-dependence of the
strength of the PDR below 10 MeV is well understood in our
model:due to the phonon coupling some of the strength in $^{48}Ca$
is simply shifted beyond 10 MeV. The predicted fragmentation of
the PDR  can be investigated in  $(e,e')$ and $( \gamma ,\gamma '
)$ experiments. Whereas the isovector dipole strength of the PDR
is small in all Ca isotopes, we find in this region surprisingly
strong isoscalar dipole states, in agreement with an $( \alpha,
\alpha ' \gamma )$ experiment.
We conclude that for the detailed understanding of the
structure of excited nuclei e.g. the PDR and GDR an approach like
the present one is absolutely necessary.
\end{abstract}

\maketitle

\newpage

 PACS: 24.30.Cz; 21.60.Ev; 27.40.+z

Keywords:
Microscopic theory,
Pygmy and giant dipole resonances,
Single-particle continuum,\\
Transition densities.

\setcounter{equation}{0}

\section{Introduction}
The pygmy dipole resonance (PDR) or "soft" electric dipole
resonances
have attracted the interest of several research groups in the past
decade \cite{palit04,kneisl96}. The PDR is  defined as a part of
the low-energy tail of the isovector dipole resonance (GDR)  and
dwells well below  the  GDR maximum.
 In phenomenological
models these resonances are described as vibrations of the excess
neutrons (neutron skin) against an inert core with N=Z
\cite{ikeda}. Therefore, one expects little strength in N=Z nuclei
and increasing strength with increasing neutron excess. In exotic
nuclei with extreme neutron excess these PDR should be especially
pronounced. As the dipole strength near the nucleon separation
energy strongly influences the r-process in astrophysics, the PDR
in such nuclei is of importance for the nuclear synthesis.

Another important reason of the interest in the PDR are  modern
experimental techniques which allow to distinguish between
1$^{-}$,2$^{+}$ and 1$^{+}$ states \cite{kneisl96,las85}. In
recent experiments new data have been obtained for several groups
of isotopes (see \cite{palit04,hartmann02,we04}) including the
stable Ca isotopes \cite{hartmann02,we04}. The latter results for
the three nuclei $^{40}Ca $,$^{ 44}Ca$ and $^{48}Ca $, which are
the bridge between light and medium weight nuclei, are especially
interesting. The experiment shows a large difference in the summed
E1 strengths of the PDR below 10 MeV for $^{40}Ca$ and $^{44}Ca$.
Surprisingly the EWSR of the PDR does not rise with the neutron
excess but is roughly the same for $^{44}Ca$ and $^{48}Ca$. This
contradicts the phenomenological models as well as microscopic RPA
calculations \cite{chamb94}.

Clearly, such an analysis should be performed within microscopic
theoretical approaches which simultaneously describes the PDR and
GDR and other nuclear structure properties. The models should not
have free fitting parameters for each nucleus in order to obtain
reliable results which allow to understand the underlying physics.
Such an analysis has been performed in ref. \cite{chamb94} for the
three Ca isotopes within the RPA based on the self-consistent
density functional method and for heavier nuclei within the
relativistic self-consistent RPA or QRPA based on the relativistic
Hartree-Bogoliubov model \cite{ring05}. In both approaches the
self-consistency ensured a reliable comparison between different
nuclei. Goriely and Khan \cite{goriely02} calculated  within the
self-consistent QRPA the E1 strength distributions
\cite{goriely02} for all nuclei with $8 \leq Z \leq 110$  between
the proton and neutron drip lines using known Skyrme forces. In
their calculation the low-lying E1 strength was located
systematically higher by some 3 MeV compared with the available
data.

In the past 15 years Kamerdzhiev, Speth, Tertychny and Tselyaev
have developed
the \emph{Extended Theory of Finite Fermi Systems(ETFFS)} which
uses the Green function formalism, is based on the RPA and
includes the single-particle continuum as well as the coupling to
the low-lying phonons. It also considers the effect of the phonons
in the ground state correlations. The authors have shown in
numerous publications (see \cite{rev} and references therein),
that this approach allows to calculate simultaneously low-lying as
well as high lying nuclear structure properties in quantitative
agreement with the experimental data. The model has been applied
to all closed shell nuclei. As far as the electric dipole states
are concerned the complete spectrum has always been  calculated,
but as the PDR are weak compared to the GDR they appear only as
small fluctuations at the lower end of the GDR.

The first calculations of the PDR within this model including
pairing have been performed for Ca \cite{we04} and Sn
 \cite{kaev04} isotopes.
 Here the characteristics of the PDR are in much better
 agreement with the data than the above mentioned self consistent
 calculations. This is due to coupling to the phonons, first of all,
 and due  to another   single-particle scheme.
 There exists a basic difference between the present approach and
  self-consistent (Q)RPA calculations e.g. \cite{chamb94,goriely02,ring05}
  as we will discuss below.

Long time ago  the conventional (1p-1h) RPA and QRPA have been
extended to include the effects of phonons. Soloviev et al.
introduced 30 years ago the \emph{Quasiparticle Phonon Model
(QPM)}\cite{soloviev,vg92}, the present model $(ETFFS)$ was first
applied nearly 15 years ago in ref.\cite{kst93,kstw93} and around
the same time Broglia et al. developed the \emph{Phonon Coupling
Model(PCM)} \cite{colo94}. However, the extended models
\cite{rev,colo94} have been used only recently  to calculate the
properties of PDR in non-magic nuclei because an additional work
had to be done to include pairing to these models . In addition to
the above mentioned results, the PDR calculations have been
performed for $^{208}Pb$ \cite{resaeva02} and for a long Sn
isotopes chain \cite{tsoneva04} within the QPM,  for
$^{120}Sn$,$^{132}Sn$ , $^{208}Pb$ (together with the giant E1
resonances)\cite{sarchi04} and for $^{18}$O,$^{20}$O,$^{22}$O
\cite{colo01} within the PCM generalized to include pairing. Very
recently, the E1 photo absorption cross sections have been
calculated in $^{116}Sn$, $^{120}Sn$ and $^{124}Sn$ with a version
of the ETFFS (see below) \cite{lt05},which also considers pairing.
So far the single-particle continuum for non-magic nuclei has been
only taken into account in the present approach (ETFFS)
\cite{we04,kaev04,lt05}.

In the recent $ETFFS$ calculations for the PDR in Ca isotopes
\cite{we04}, several approximations (for details see
\cite{kaev04}) have been made which were responsible that the
theoretical results agreed only qualitatively with the data. The
aim of the present investigation is threefold: firstly, we repeat
the previous calculations without the approximations and apply for
$^{44}Ca$ the generalized version of the $ETFFS$ which includes
the pairing effects, secondly we clarify the role of the continuum
for PDR and GDR, which has attracted a great interest now
\cite{richter05}, and finally we investigate in detail the
microscopic structure of the PDR, especially the role of the
isoscalar dipole components. The result of the present calculation
agrees well with the experimental
 data \cite{hartmann02,we04} and it explains in a natural way
 the somewhat  surprising result that the strength of the PDR does
 (seemingly) not scale with the neutron excess.

\section{Method}

We have used the generalized version of the ETFFS with pairing
which differs from the one used in \cite{we04,kaev04} by  taking
into account the self-energy and the induced ph- and pp-
interaction graphs on the same footing, i.e. by a consistent
summations  of the so-called g$^{2}$ diagrams where g is the
amplitude for  low-lying phonon creation. The formalism developed
for nuclei with pairing, is  called \emph{ Quasiparticle Time
Blocking Approximation (QTBA)}, and is described in detail in
\cite{tselyaevQTBA} and \cite{lt05}.(However, the ground state
correlations induced by the phonon coupling have not been included
to the present calculations ). The single-particle continuum has
been included on the RPA level where it is taken into account
correctly  within  our Green function technique in the coordinate
representation. For details see \cite{rev}. Therefore in our
approach we consider the three mechanisms which create the width
of giant resonance, namely (I) the Landau damping ((Q)RPA
configurations),(II) the escape width (the single-particle
continuum) and (III) the spreading width (phonon coupling, or
complex configurations). As in all our previous calculations
within the ETFFS, we include the most collective low-lying
phonons, namely 3$^{-}_{1}$, 5$^{-}_{1}$ phonons for $^{40}$Ca,
and 2$^{+}_{1}$,3$^{-}_{1}$,5$^{-}_{1}$ phonons for $^{44}$Ca and
$^{48}$Ca. The collective phonons have been microscopically
calculated within the RPA for $^{40}$Ca and $^{48}$Ca. In the case
of $^{44}$Ca we used the corresponding QRPA with pairing. The
standard BCS equation have been applied using the
particle-particle Landau-Migdal forces \cite{sap}.
 In the present and in  our previous calculations we used
 the special "forced consistency" procedure \cite{kl98}
 to obtain the energy of the spurious E1 state to be exactly equal
 to zero without the procedure of fitting force parameters.

As in our previous calculations we used the  effective
particle-hole Landau-Migdal interaction:

\begin{eqnarray}
F({\bf r},{\bf r'}) = C_{0}[f(r) + f'( r) {\bf {\bftau}_{1}}
\cdot{\bf {\bftau}_{2}} +\nonumber \\
 (g + g' {\bf {\bftau}_{1}} \cdot {\bf
{\bftau}_{2}} ) {\bf {\bfsigma}_{1}} \cdot {\bf {\bfsigma}_{2}}
] \;{\delta} ({\bf r} - {\bf r'}),
\end{eqnarray}
with the conventional interpolation formula, for example, for the
parameter f
\begin{equation}
f(r) = f_{ex} + (f_{in} - f_{ex})\rho_{0}(r)/\rho_{0}(0)
\end{equation}
and similarly for the other $r$-dependent parameters.
 Here $\rho_{0}(r)$ is the
density distribution of the ground state of the nucleus under
consideration and $f_{in}$ and $f_{ex}$ are the force parameters
inside and outside of the nucleus. The standard  values of the
parameters, which have been already used for all the nuclei under
consideration \cite{rev,we04,kaev04},  are as follows
\begin{eqnarray}
f_{in}  =  - 0.002,\; f_{ex} =-1.4,\;
f_{ex}^{\prime}  =  2.30, \;
f_{in}^{\prime}  =  0.76, \nonumber \\ \nonumber \\
g  =  0.05,\;
g^{\prime}  =  0.96, \;
C_{0}  =  300\;{\rm MeV fm^{3}}.
\end{eqnarray}
For the nuclear density $\rho_{0}(r)$ in the interpolation formula
 we chose the theoretical  ground state density distribution
of the corresponding nucleus,
\begin{equation}
\rho_{0}(r) = \sum_{\epsilon_{i} \leq \epsilon_{F}} \frac{1}{4\pi}
(2j_{i} + 1)   R^{2}_{i} (r),
\end{equation}
which is more consistent than the previously used Woods-Saxon
 distribution. For that reason we had to readjust $f_{ex}$ and
 obtained the value of $f_{ex}$ = -1.4 used \cite{kst97}.
Here  $R_{i}(r)$ are the single-particle radial wave functions of
the  single-particle model used. For other details of the
calculations, see \cite{rev,we04,kaev04}.
\section{ PDR Results}
\begin{figure}[ht]
\centerline{\includegraphics[width=1.\linewidth,height=0.8\linewidth,angle=0]{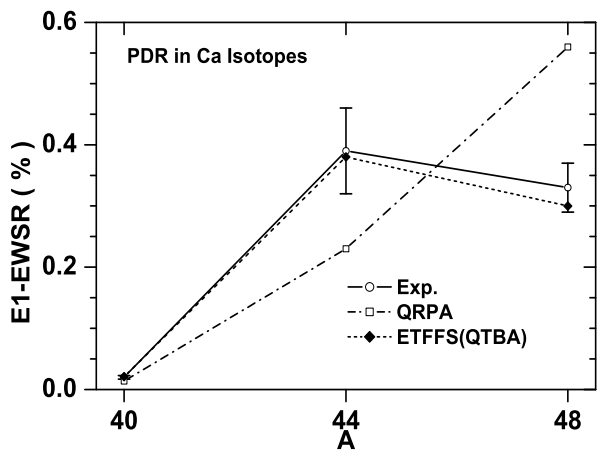}}
\caption{The energy-weighted electric dipole  strength integrated
up to 10 MeV is shown for the isotopes $^{40}$Ca, $^{44}$Ca,  and
$^{48}$Ca. Dashed: experimental data taken from \cite{we04};
dash-dotted: quasiparticle random phase approximation, full:
extended theory of finite Fermi systems in the quasiparticle time
blocking approximation (including pairing for $^{44}$Ca )}
\end{figure}

The main results of our  calculations for the gross properties PDR
are given in Fig.1 and  Table~\ref{tab:t1}.
\begin{table*}
\caption{Properties of the pygmy dipole resonances in Ca isotopes
in the (5 - 10) MeV interval.\\(The experimental data  taken from
\cite{we04}) } \label{tab:t1}
\begin{center}
\begin{tabular}{|c|c|c|c|c|}
\hline
                       &             & $^{40}$Ca & $^{44}$Ca  & $^{48}$Ca  \\
\hline
$\sum_i B_i(E1)$$ \uparrow$, & (Q)RPA      & 2.51 & 45.6 & 97.4 \\
$10^3  e^2$fm$^2$    & ETFFS(QTBA) & 3.67 & 81.0 & 66.2 \\
        & Exp.         & 5.1 $\pm$ 0.8 & 92.2 $\pm$ 15.6 & 68.7 $\pm$ 7.5 \\
\hline
$\sum_i E_i \, B_i(E1)$$ \uparrow$, & (Q)RPA & 21.48 & 374.9 & 897.3  \\
keV$e^2$fm$^2$ & ETFFS(QTBA) & 31.86 & 611.0 & 527.7  \\
    & Exp. & 35 $\pm$ 5 & 629  $\pm$ 107 & 570 $\pm$ 62 \\
\hline
$\sum_i E_i \, B_i(E1) \!\!\uparrow$, & (Q)RPA & 0.014 & 0.23 & 0.56 \\
\% EWSR & ETFFS(QTBA) & 0.021 & 0.38 & 0.30 \\
                      & Exp. & 0.020 $\pm$ 0.003 & 0.39 $\pm$ 0.07 & 0.33 $\pm$ 0.04\\
\hline
               & (Q)RPA & 8.84 & 8.22 & 9.30 \\
$\bar{E}$, MeV & ETFFS(QTBA) & 8.68 & 7.54 & 7.97 \\
               & Exp. & 6.80 & 7.10 & 8.40 \\
\hline
\end{tabular}
\end{center}
\end{table*}
Within the experimental errors, our theoretical results agree with
the data as shown in $TableI$. This holds for the total B(E1)
strength, EWSR and the EWSR share in the energy interval up to 10
MeV. As for the
 PDR mean energies, which  were defined as $\overline{E}
=\Sigma E_{i}B_{i}(E1)/\Sigma B_{i}(E1)$,
the agreement is rather good for $^{44}Ca$ and $^{48}Ca$ but the
theoretical result 8.68 MeV is too high
 in $^{40}Ca$. In all the
cases the results of the extended theory, i.e. inclusion of phonon
coupling and single-particle continuum, give rise to a
considerable improvement compared to the (Q)RPA results.
 and (especially for
$^{44}$Ca) our results of ref. \cite{we04}. The latter
 is due to an improvement of the theory but even more importantly
to more careful calculations of the single-particle levels and of
the (Q)RPA phonons, including the use of eq.(2.4) for them.

  In Fig.1 we compare the
theoretical EWSR of $^{40}Ca$,$^{44}Ca$ and $^{48}Ca$ for the
isovector dipole strength up to 10 MeV with the experimental
data. The dash-dotted line is the (Q)RPA result which rises with
neutron excess in agreement with the phenomenological model. The
experiment shows only a strong increase for $^{44}Ca$ compared to
$^{40}Ca$ but the strength of $^{48}Ca$ and $^{44}Ca$ does not
change within the experimental errors. The ETFFS result reproduces
this somewhat surprising result.
\begin{figure}[ht]
\centerline{\includegraphics[width=1.\linewidth,height=0.8\linewidth,angle=0]{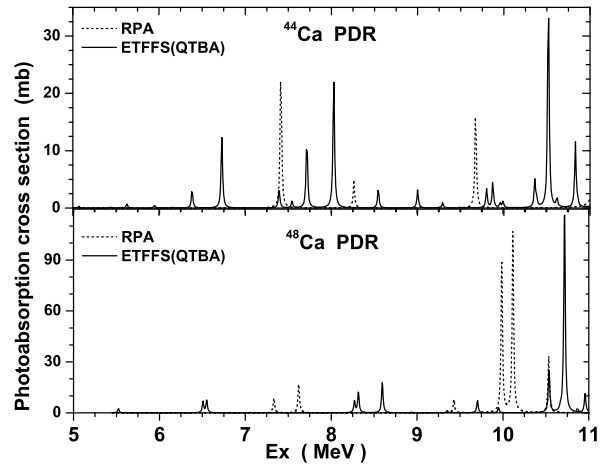}}
\caption{Pygmy resonances in $^{44}$Ca and $^{48}Ca$ up to 11 MeV}
\end{figure}

The explanation is given in Fig.2 where we compare the theoretical
PDR-spectrum of $^{44}Ca$ and $^{48}Ca$. Whereas the (Q)RPA result
is in both cases around 10 MeV and below, the ETFFS result differs
in the two nuclei. Due to the phonon splitting
an appreciable fraction of the strength in $^{48}Ca$ is shifted
to about 11 MeV and has therefore not be measured in the cited
experiment. If one adds this part ( the 10 MeV limit is somewhat
arbitrary ) the total PDR strength increase also in ETFFS with the
 neutron excess. One also realizes that the PDR for the open shell
 nucleus $^{44}Ca$ is somewhat lower compared to the PDR in the closed
 shell nucleus $^{48}Ca$.
\begin{figure}[ht]
\centerline{\includegraphics[width=1.\linewidth,height=0.8\linewidth,angle=0]{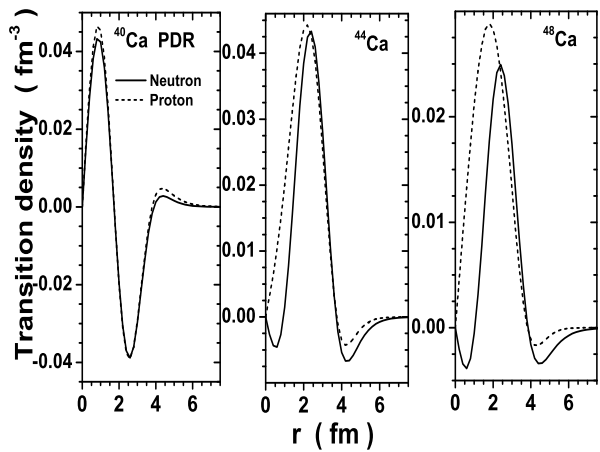}}
\caption{Transition densities of the electric dipole strength in the (5-10)MeV interval}
\end{figure}

In Fig.3 the transition densities for protons and neutrons of the
electric dipole isovector strength up to 10 MeV are  shown for the
three Ca isotopes. In $^{40}Ca$ the density is of purely isoscalar
nature. With increasing neutron excess the neutron density is
slightly shifted to a larger radius, indicating some an isovector
admixture, but the isoscalar character remains. This is easily
understood in the framework of  microscopic models. The isovector
ph-interaction is repulsive and shifts the low-lying isovector
strength into the region of the GDR. On the other hand, the
isoscalar ph-interaction is attractive and shifts some of the high
lying isoscalar dipole strength to lower energies and gives rise
to some isoscalar collectivity below 10 MeV in contrast to the
isovector case.
\begin{figure}[ht]
\centerline{\includegraphics[width=1.\linewidth,height=0.8\linewidth,angle=0]{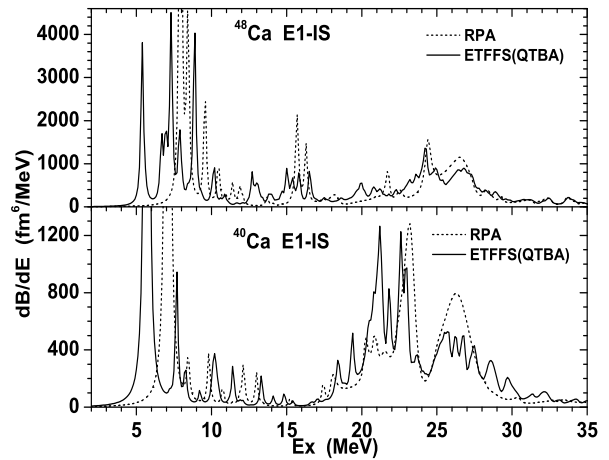}}
\caption{Isoscalar electric dipole strength in $Ca^{40}$ and $Ca^{48}$}
\end{figure}

In Fig.4 the isoscalar strength distributions (with the operator
${\bf r}^{3}Y_{10}$) in $^{40}Ca$ and $^{48}Ca$ calculated within
the ETFFS and RPA are shown. In $^{40}Ca$ the influence of the
phonons are small compared to $^{48}Ca$ especially for the low
energy spectrum. Whereas in $^{40}Ca$ the phonons only give rise
to an energy shift one observes in $^{48}Ca$ also some
fragmentation of the low lying strength. This fragmentation is due
to the low-lying $2^+$ state which can be coupled to 1p1h 1$^-$
states to produce the necessary 1$^-$ complex configurations. The
corresponding states  do not exist in $^{40}Ca$. The two peaks in
$^{40}Ca$ at 5.8 MeV and 7.8 MeV are in an agreement with the
experimental finding \cite{hvw01}.

 \section{The giant resonance results}
\begin{figure}[ht]
\centerline{\includegraphics[width=1.\linewidth,height=0.8\linewidth,angle=0]{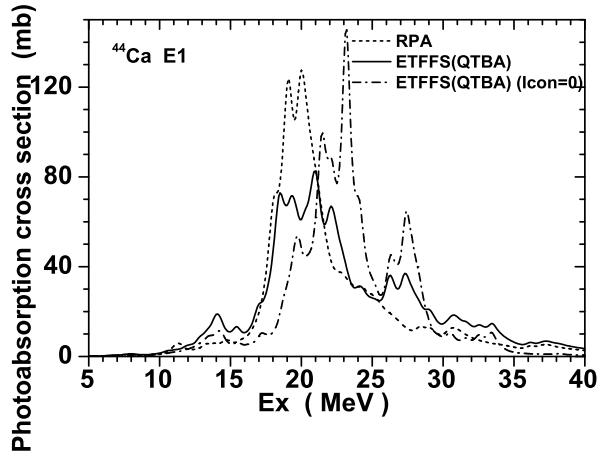}}
\caption{Giant electric dipole resonance in $^{44}Ca$ calculated
in three different approximations:(I)ETFFS(QTBA) (full line), (II)
QRPA (dash-dotted); (III) ETFFS(QTBA) without continuum
(dash-dotted)}
\end{figure}
For consistency we also calculated the giant isovector resonances in the three calcium isotopes. Results are presented in
Table~\ref{tab:t2}
and in Figs.5 and 6. In Table 2 the results obtained in three different
 approximations are shown: (Q)RPA calculations with the
 single-particle continuum as well as
 the  ETFFS(QTBA)
 calculations with and without the continuum.
 The theoretical values are extracted from the calculated strength
 distribution under the condition that the  three lowest
  energy-weighted
moments of the Lorentzian and the theoretical curve should
coincide in the (10-40) MeV interval \cite{kst93,tselyaev00}.
\begin{table}
\caption{Properties of the isovector giant dipole resonances
\protect \\ in Ca isotopes }
\label{tab:t2}
\begin{center}
\begin{tabular}{|c|c|c|c|c|}
\hline
\multicolumn{2}{|c|}{}  & $^{40}$Ca & $^{44}$Ca  & $^{48}$Ca  \\
\hline
             & (Q)RPA       & 102.3 & 106.6 & 91.2 \\
\cline{2-5}
\% EWSR     & ETFFS(QTBA)-c$^*$  & 100.4 & 108.6 & 91.4 \\
\cline{2-5}
        & ETFFS(QTA)B  & 103.7 & 110.5 & 93.0 \\
\cline{2-5}
       & Exp.               & 106.3 &       & 119.5  \\
\hline
       & (Q)RPA             & 21.0  & 20.4  & 16.8  \\
\cline{2-5}
$\bar{E}$, MeV & ETFFS(QTBA)-c       & 22.7 &  23.1  & 19.9  \\
\cline{2-5}
       & ETFFS(QTBA) & 20.9 & 21.2  & 18.4  \\
\cline{2-5}
       & Exp.               & 20.0 &        & 19.6  \\
\hline
       & (Q)RPA             & 137.2 & 146.6 & 170.6 \\
\cline{2-5}
$\sigma_{\mbox{max}}$, mb & ETFFS(QTBA)-c       & 137.2 & 129.2 & 113.7  \\
\cline{2-5}
       & ETFFS(QTBA) & 75.6 & 80.9 & 85.4  \\
\cline{2-5}
       & Exp.               & 95.0  &       & 102.7  \\
\hline
       & (Q)RPA             & 2.9   & 3.2   & 2.6  \\
\cline{2-5}
$\Gamma$, MeV & ETFFS(QTBA)-c       & 3.2   & 3.8   & 4.1  \\
\cline{2-5}
       & ETFFS(QTBA) & 6.2   & 6.4   & 5.7  \\
\cline{2-5}
       & Exp.               & 5.0   &     & 7.1  \\
\hline
\end{tabular}
\\$^*$ -c stands for calculations without continuum.
\end{center}
\end{table}
The averaging  parameter $\Delta $ which was used  in addition to
the single-particle continuum the giant resonance  calculation was
400 keV. The differences between  the (Q)RPA and full ETFFS(QTBA)
are most striking for the width $\Gamma$ and the maximum of the
photo absorption cross section $\sigma_{max}$, where the results
differ nearly by a factor of two. We obtain good agreement with
the experiments only within our full model, i.e. including the
phonons and the single-particle continuum. Here we want to stress
that we calculate GDR and the low-lying states simultaneously
within the same configuration space and with the same forces.
\begin{figure}[ht]
\centerline{\includegraphics[width=1.\linewidth,height=0.8\linewidth,angle=0]{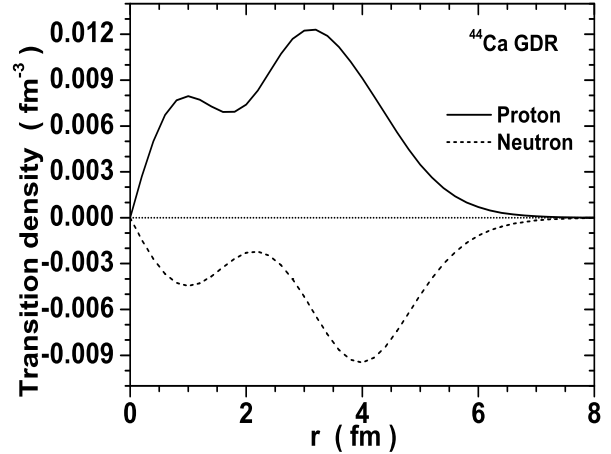}}
\caption{Transition densities for $^{44}$Ca; the isovector
electric dipole strength is integrated in the (15-30) MeV
interval}
\end{figure}

As compared with our earlier results for $^{40}Ca$ and $^{48}Ca$
 \cite{kst93,kst97}, here
 the ground state correlations induced by the phonon coupling
were not taken into account and we have chosen a different energy
interval for the summation. In addition, a different
 definition for the mean energy and another "refining" procedure
 \cite{tselyaevQTBA} have been used.
Like in the previous calculations of giant multipole resonances
in closed shell nuclei \cite{rev}
the inclusion of phonon coupling , i.e. the calculation
within the full ETFFS (QTBA), increases the widths by a factor of 2
 and (except for $^{40}$Ca) shifts the mean energies by 1.0-1.5 MeV
 to higher energy.

 In Fig.5 the results for the GDR in $^{44}Ca$ are  shown,
calculated again within the three approximations. In the full
model an appreciable fraction of the strength is shifted to much
higher energies compared to the QRPA result which give rise to an
asymmetric shape. Finally in Fig.6 the transition density for the
GDR in $^{44}Ca$ is plotted. It has the expected isovector shape.

In order to study {\it the role of the single-particle continuum}
we have  calculated the   giant resonance (see Table 2 and Fig.5)
and PDR characteristics without the continuum. It should be
expected that the role of the single-particle continuum is
important in light nuclei. It is of special interest to consider
the question within the same calculational scheme for the Ca
isotopes, including  the non-magic $^{44}$Ca, because there is no
information about the role of the continuum in the approaches of
similar kind in non-magic nuclei. One can see from Table 2 and
Fig.5 a very noticeable influence of the continuum for the
integral characteristics in  all the Ca isotopes as well as for
the strength distribution. Moreover, the continuum's influence is
noticeable even for the PDR : our full calculations have given
0.38\% of the EWSR in the (5.0-10.0) MeV interval for $^{44}$Ca
whereas the calculation without continuum gave 0.19\%. The
exclusion of the continuum also results in a considerable
redistribution of the PDR strength. These features can be studied
in experiments with electrons and gamma-rays.

\section{Summary}

 We have investigated the low-lying (PDR)
 and high-lying (GDR) electric dipole states
in the magic $^{40}$Ca and $^{48}$Ca and non-magic $^{44}$Ca
within the ETFFS(QTBA) which, in addition to the standard (Q)RPA,
takes into account the single-particle continuum and  phonon
coupling. In the present approach we corrected for the double
counting due to  the phonons in our generalized propagator. Good
agreement with experiment available for integral characteristics
has been found.

The agreement for the PDR is much better as compared with the
previous ETFFS calculations \cite{we04}.
 The PDR, i.e. the
isovector dipole strength below 10 MeV, is less than 1 percent of
the energy weighted sum rule because the strongly repulsive
isovector force shift the strength into GDR region.
 In order to investigate the structure of the PDR further
we calculated the corresponding transition densities. It turned
out that these densities are isoscalar with some isovector
admixture in $^{44}$Ca and $^{48}$Ca which  is due to the strongly
attractive
 isoscalar interaction which shifts the high-lying isoscalar strength
 below 10 MeV.

In $^{40}$Ca  we obtain two prominent isoscalar dipole states
(induced by the external field ${\bf r}^{3}Y_{10}$) at 5.8 MeV and
7.8 MeV. The predicted isoscalar strength can be detected in
${(\alpha,\alpha'\gamma)}$ experiments, see \cite{zielges05}.

 We have shown that the role of the single-particle continuum is
very noticeable  for the GDR and quantatively important for the
PDR in the Ca isotopes considered. The same should be  true for
lighter nuclei too.
 Thus, taking into account the new effects as
compared with the standard RPA or QRPA , i.e. the phonon coupling
and  single-particle continuum , is necessary to explain the
properties of both these resonances.

\section{Acknowledgments}
The authors are very thankful to N.
Luytorovich for permanent collaboration in the preparation of this
article and to
 A.Zilges for the information
about the experiments \cite{zielges05}. S.Kamerdzhiev thanks
 the Institute
for Nuclear Theory (Seattle, USA) for partial support during the
participation in the Program INT-05-3 which was very useful for
this work. The work was supported in part by the DFG and RFBR
grants Nos.GZ:432RUS113/806/0-1 and 05-02-04005 and by the INTAS
grant No.03-54-6545.

\end{document}